\documentclass[notitlepage,a4,10.5pt,twocolumn]{article}

\usepackage{graphicx}

\usepackage{amsmath}%
\usepackage{amsfonts}%
\usepackage{amssymb}%
\usepackage{mathrsfs}
\usepackage{amsthm}
\usepackage{float}

\usepackage{authblk}

\usepackage[left=2.5cm,right=2.5cm,top=3cm,bottom=3cm]{geometry}

\newcommand{\Norm}[1]{\left\lvert\left\lvert #1 \right\rvert\right\rvert}

\newcommand{\bol}{\boldsymbol}

\newcommand{\abs}[1]{\left\lvert{#1}\right\rvert}
\newcommand{\w}{\wedge}
\newcommand{\lr}[1]{\left({#1}\right)}

\newcommand{\mf}{\mathfrak}
\newcommand{\p}{\partial}

\newcommand{\ti}[1]{\textit{#1}}
\newcommand{\tb}[1]{\textbf{#1}}

\begin{document}

\title{Symmetric Ideal Magnetofluidostatic Equilibria with Non-Vanishing Pressure Gradients in Asymmetric Confinement Vessels}
\author[1]{Naoki Sato} 
\affil[1]{Graduate School of Frontier Sciences, \protect\\ The University of Tokyo, Kashiwa, Chiba 277-8561, Japan \protect\\ Email: sato\_naoki@edu.k.u-tokyo.ac.jp}
\date{\today}
\setcounter{Maxaffil}{0}
\renewcommand\Affilfont{\itshape\small}

\twocolumn[
  \begin{@twocolumnfalse}
    \maketitle
    \begin{abstract}
    We study the possibility of constructing steady magnetic fields satisfying the force balance equation of ideal magnetohydrodynamics with tangential boundary conditions in asymmetric confinement vessels, i.e. bounded regions that are not invariant under continuous Euclidean isometries (translations, rotations, or their combination). This problem is often encountered in the design 
of next-generation fusion reactors. We show that such configurations are possible
if one relaxes the standard assumption that the vessel boundary corresponds to a pressure isosurface. 
We exhibit a smooth solution that possesses an Euclidean symmetry and yet solves the boundary value problem in an asymmetric ellipsoidal domain while sustaining a non-vanishing pressure gradient. 
This result provides a definitive answer to the problem of existence of regular ideal magnetofluidostatic equilibria in asymmetric bounded domains. The question remains open whether regular asymmetric solutions of the boundary value problem exist.     
\end{abstract}
\vspace{5mm}
\end{@twocolumnfalse}
  ]

\section{Introduction}
In a static ideal magnetofluid equilibrium \cite{Grad58}, the Lorentz force is 
counterbalanced by a pressure gradient:
\begin{subequations}\label{IMHD}
\begin{align}
\lr{\nabla\w\bol{B}}\w\bol{B}=&\nabla P,\label{IMHD1}\\
\nabla\cdot\bol{B}=&0.
\end{align}
\end{subequations}
Here, $\bol{B}=\lr{B_x,B_y,B_z}^T$ is the magnetic field, and $P$ the pressure.
Usually, equation \eqref{IMHD} is solved within a bounded region $\Omega\subset\mathbb{R}^3$ with tangential boundary conditions:
\begin{equation}
\bol{B}\cdot\bol{n}=0~~~~{\rm on}~~\p\Omega,\label{BC}
\end{equation}
where $\bol{n}$ is the unit outward normal to the boundary $\p\Omega$ (typically, a two dimensional toroidal surface). 

At present, a rigorous mathematical treatment of system \eqref{IMHD}
with boundary conditions \eqref{BC} is not available.
This difficulty stems from the mixed nature
of these equations: they define a nonlinear twice hyperbolic twice elliptic 
first order system of PDEs for the variables $B_x$, $B_y$, $B_z$, and $P$ (see \cite{Yoshida90,SatoPPCF} on this point).  
For this reason, the existence of solutions with appropriate regularity and symmetry properties represents an open mathematical problem with  both practical and theoretical implications \cite{LoSurdo}.

On the practical side, it is thought that certain asymmetric field configurations 
(here, a symmetry is an invariance under a continuous Euclidean isometry) 
may provide a substantial advantage with regard to particle confinement
in next-generation nuclear fusion reactors (stellarators) when compared with 
classical toroidal devices (tokamaks) \cite{Hudson18}. 
The optimization of particle losses due to drift motion
is usually achieved by imposing a `quasisymmetry' condition
on the magnetic field. 
A quasisymmetry is an asymmetric field configuration such that
the field strength is invariant along a solenoidal direction in space.
In practice, asymmetric solutions are approximated by 
using near-axis expansions that, however, lead to overdetermined systems
of equations \cite{Garren91,Garren91_2} 
or series representations whose convergence is not guaranteed \cite{Weitzner16}.
For this reason, the existence of quasisymmetric solutions remains elusive.
In fact, even without imposing the quasisymmetry condition, 
the existence of sufficiently regular solutions capable of
sustaining a non-vanishing pressure gradient in an asymmetric domain
is still an unsolved problem that we wish to address in the present paper.
 
On the other hand, revealing the mathematical structure
of system \eqref{IMHD}, \eqref{BC} may provide insight into the rigourous 
treatment of elliptic-hyperbolic systems of PDEs.
Moreover, magnetofluidostatic equilibria are mathematically equivalent
to steady solutions of the incompressible ideal Euler equations of fluid dynamics
with constant fluid density \cite{Moffatt14}. For this reason, the theory has implications for 
the properties of solutions to the incompressible Euler and Navier-Stokes equations. 

The admissible field topologies associated with solutions of system \eqref{IMHD}, \eqref{BC} such that
the magnetic field and the electric current are not everywhere collinear  
are described by Arnold's structure theorem \cite{Arnold66,Enciso14},
according to which the domain $\Omega$ can be divided into a finite number of subdomains where $\bol{B}$ 
is either tangent to two dimensional tori or cylindrical surfaces.  
The challenge posed by system \eqref{IMHD}, \eqref{BC}
can be understood from a geometrical/topological standpoint, 
a functional/variational standpoint, or a Lie-symmetry perspective. 
To see this, it is convenient to approach the problem
by introducing Clebsch potentials $\psi$ and $\theta$ such that
\begin{equation}
\bol{B}=\nabla\psi\w\nabla\theta.\label{Clebsch}
\end{equation}
Recall that this representation is always valid in any sufficiently
small neighborhood $U\subset\Omega$ of a chosen point of interest due to
the Lie-Darboux theorem \cite{deLeon,Arnold89}, but it is an incomplete parametrization when the whole region $\Omega$ is considered (i.e. more Clebsch parameters may be needed to represent $\bol{B}$ in the whole $\Omega$, see \cite{Yoshida09}). 
Due to the solenoidal nature of $\bol{B}$ and $\nabla\w\bol{B}$, and the properties $\bol{B}\cdot\nabla P=0$, $\nabla\w\bol{B}\cdot\nabla P=0$ arising from \eqref{IMHD}, if $\nabla P\neq\bol{0}$ we can always construct a local
coordinate system $\lr{x^1,x^2,x^3}$ such that $x^3=P$ and
\begin{equation}
\bol{B}=\nabla x^2\w\nabla x^3,~~~~\nabla\w\bol{B}=\nabla x^3\w\nabla x^1.
\end{equation}
Then, equation \eqref{IMHD} is equivalent to the following conditions
on the tangent vectors $\lr{\p_1,\p_2,\p_3}$ of the coordinate system $\lr{x^1,x^2,x^3}$:
\begin{subequations}\label{IMHD2}
\begin{align}
\nabla\w \p_1=&\p_2,\label{d12}\\ 
\p_1\cdot\p_2\w\p_3=&-1.
\end{align}
\end{subequations}
Hence, finding a solution of equation \eqref{IMHD} with a non-vanishing pressure gradient is locally equivalent to the geometrical statement 
that there exists a coordinate system such that
the curl of the first tangent vector gives the second tangent vector
and the Jacobian determinant of the transformation is minus unity. 
Equation \eqref{d12} represents a stringent geometrical condition 
that restricts the topology of nontrivial solutions. 

In terms of the covariant metric tensor $g_{ij}=\p_i\cdot\p_j$, one has $\p_1=g_{1j}\nabla x^j$.  
Then, \eqref{IMHD2} translates into the following requirements on the coefficients $g_{ij}$: 
\begin{subequations}\label{IMHD3}
\begin{align}
\frac{\p g_{13}}{\p x^1}-\frac{\p g_{11}}{\p x^3}=&1,\\
\frac{\p g_{13}}{\p x^2}-\frac{\p g_{12}}{\p x^3}=&0,\\
\frac{\p g_{12}}{\p x^1}-\frac{\p g_{11}}{\p x^2}=&0,\\
\epsilon^{ijk}g_{1i}g_{2j}g_{3k}=&1.
\end{align}
\end{subequations}

Next, observe that a direct substitution of the parametrization $\bol{B}=\nabla \psi\w\nabla \theta$ into \eqref{IMHD} gives
\begin{equation}
\left[\nabla\w\lr{\nabla \psi\w\nabla \theta}\right]\w\lr{\nabla \psi\w \nabla \theta}=P_{\psi}\nabla\psi+P_{\theta}\nabla\theta.\label{psith} 
\end{equation}
Here, we used the fact that $\nabla P\cdot\bol{B}=0$ implies $P=P\lr{\psi,\theta}$. 
By application of standard vector identities, equation \eqref{psith} can be reduced to 
the following system of coupled nonlinear second order PDEs for the
Clebsch potentials $\psi$ and $\theta$:
\begin{subequations}\label{IMHD4}
\begin{align}
\nabla\cdot\left[\nabla \theta\w\lr{\nabla \psi\w\nabla \theta}\right]=&-P_{\psi},\\
\nabla\cdot\left[\nabla \psi\w\lr{\nabla \theta\w\nabla \psi}\right]=&-P_{\theta}.
\end{align}
\end{subequations}
Therefore, system \eqref{IMHD} is locally equivalent to system \eqref{IMHD4}. 
System \eqref{IMHD4} 
can be formally obtained by extremizing the the energy functional
\begin{equation}
H=\int_{\Omega}\lr{\frac{1}{2}\abs{\nabla\psi\w\nabla\theta}^2-P\lr{\psi,\theta}}dV.
\end{equation}
However, the functional $H$ is not coercive \cite{Struwe}, 
meaning that one can increase the $L^2\lr{\Omega}$ norms $\Norm{\nabla\psi}$, $\Norm{\nabla\theta}$
while keeping $H$ arbitrarily small. 
This is the reason why standard methods of functional analysis cannot be applied in a straightforward manner to 
determine the existence of solutions,   
and the availability of a minimum of $H$ (i.e. a solution of \eqref{IMHD4}) is not guaranteed in general.  

Finally, system \eqref{IMHD} can also be interpreted as a symmetry statement. 
Indeed, taking the curl of \eqref{IMHD1} and recalling the vector identity
\begin{equation}
\nabla\w\left[\bol{B}\w\lr{\nabla\w\bol{B}}\right]=\mf{L}_{\nabla\w\bol{B}}\bol{B},
\end{equation}
where
\begin{equation}
\mf{L}_{\nabla\w\bol{B}}\bol{B}=
\lr{\nabla\w\bol{B}\cdot\nabla}\bol{B}-\lr{\bol{B}\cdot\nabla}\nabla\w\bol{B}, 
\end{equation}
is the Lie derivative of vector fields, equation \eqref{IMHD1} reads as
\begin{equation}
\mf{L}_{\nabla\w\bol{B}}\bol{B}=\bol{0}.
\end{equation} 
This equation implies that solutions $\bol{B}$ of system \eqref{IMHD}
are Lie-invariant along $\nabla\w\bol{B}$, i.e. $\nabla\w\bol{B}$ must be a direction of symmetry.  

Despite these difficulties, several classes of solutions of system \eqref{IMHD} are known and
can be divided in three main categories: harmonic solutions (class $h$), Beltrami field solutions (class $b$), 
and non-vanishing pressure gradient solutions (class $p$). 
A harmonic solution of system \eqref{IMHD}, \eqref{BC} 
is a current-free magnetic field configuration $\nabla\w\bol{B}=\bol{0}$ such that $\bol{B}=\nabla\phi$ for some potential $\phi$. 
Notice that for a harmonic solution the pressure $P$ is constant.
Harmonic solutions can be obtained by solving the Neumann boundary value problem for Laplace's equation:
\begin{equation}
\Delta\phi=0~~~~{\rm in}~~\Omega,~~~~\nabla\phi\cdot\bol{n}=0~~~~{\rm on}~~\p\Omega.\label{har}
\end{equation}  
We remark that the only smooth solutions of \eqref{har} are $\phi=C$, with $C\in\mathbb{R}$ \cite{Mikhailov}.
Non-constant solutions arise when $\phi$ is an angle (multivalued) variable and the domain $\Omega$ is multiply connected \cite{Yoshida09,Kodaira49}. 
For example, the toroidal angle $\theta=\arctan\lr{y/x}$ 
gives a simple symmetric harmonic solution in an axially symmetric torus, $\bol{B}=\nabla\theta=\lr{x\nabla y-y\nabla x}/(x^2+y^2)$. 
While this magnetic field is smooth in any toroidal region not containing the $z$ axis, the angle $\theta$ is multivalued, 
being an inverse trigonometric function. 
Another example of a harmonic solution in a toroidal region with Euclidean (flat) metric
can be found in \cite{Weitzner20}.  

Beltrami field solutions are defined by the force-free condition $\lr{\nabla\w\bol{B}}\w\bol{B}=\nabla P=\bol{0}$,
which implies that $\nabla\w\bol{B}=\hat{h}\bol{B}$ for some proportionality coefficient $\hat{h}$. 
Beltrami fields can be further divided in those that possess a constant proportionality factor $\hat{h}\in\mathbb{R}$ (class b1),
and those in which $\hat{h}$ is a function of the spatial coordinates $\hat{h}=\hat{h}\lr{\bol{x}}$ (class b2).
A classical theorem of Yoshida and Giga \cite{Yoshida90_2} guarantees the
existence of solutions to the boundary value problem
\begin{equation}
\nabla\w\bol{B}=\hat{h}\bol{B}~~~~{\rm in}~~\Omega,~~~~\bol{B}\cdot\bol{n}=0~~~~{\rm on}~~\p\Omega,
\end{equation}
for all $\hat{h}\in\mathbb{R}$ provided that $\Omega$ is multiply connected (the bounded region $\Omega$ must have at least one hole). 
We remark that such solutions  
belong to the standard Sobolev space $H^1$. 
An explicit example of Beltrami field with constant proportionality factor is the ABC flow \cite{Dombre86}, 
which is the prototype of flows with chaotic behavior. 
However, when $\hat{h}$ is not constant, equation \eqref{IMHD} does not have solution
for most choices of $\hat{h}$ (see \cite{Enciso16}). 
A method to construct Beltrami fields with non-constant $\hat{h}$ can be found in \cite{SatoPhysD1}. 
 
Finally, non-vanishing pressure gradient solutions satisfy $\nabla P\neq\bol{0}$ in some subset of their domain.
Existence results for this class rely on the notion of symmetric solution:
given a coordinate system $\lr{x^1,x^2,x^3}$, if the quantities $B^i$, $P$, and $g_{ij}$ with $i,j=1,2,3$
are independent of $x^3$, then system \eqref{IMHD} can be reduced to a solvable single nonlinear second order elliptic
PDE for the stream function, the Grad-Shafranov equation \cite{Ed1,Ed2}. 
It is customary to solve the Grad-Shafranov equation by demanding that the stream function is constant
on the boundary $\p\Omega$ \cite{Kruskal58,Grad67}. Since, in this setting, the pressure is a function of the stream function, it follows that
solutions of the Grad-Shafranov equation and the boundary $\p\Omega$ possess the same symmetry. 
However, as it will be shown in the present paper, a symmetric solution does not imply a symmetric boundary in general.

If one is willing to allow discontinuities,  
class $p$ solutions of \eqref{IMHD}, \eqref{BC} can be constructed in asymmetric tori \cite{Bruno96}, 
and class $p$ asymmetric solutions of the boundary value problem \eqref{IMHD}, \eqref{BC} can be obtained
in arbitrary domains \cite{SatoPPCF}. 
Nonetheless, the existence of class $p$ regular solutions in asymmetric bounded domains,
and the existence of class $p$ regular asymmetric solutions in bounded domains remain open issues. 

Table \ref{tab1} summarizes the properties of known regular solutions. 
Explicit expressions can be found in the appendix. 
\begin{table}
\begin{center}
\begin{tabular}{ |c|c|c| }
\hline
Class & Solutions of \eqref{IMHD} & Solutions of \eqref{IMHD}, \eqref{BC}\\
\hline
$h$ & $SC^{\infty}$, $AC^{\infty}$ & $SC^\infty\Omega_S$\\ 
$b1$ & $SC^{\infty}$, $AC^{\infty}$ &  $H^1\Omega_S$\cite{Yoshida90_2}, $H^1\Omega_A$\cite{Yoshida90_2}\\
$b2$ & $SC^{\infty}$, $AC^{\infty}$\cite{SatoPPCF} &  \\
$p$ & $SC^{\infty}$, $AC^{\infty}$\cite{SatoPPCF} & $SC^\infty\Omega_S$, $SC^{\infty}\Omega_A^\ast$\\
\hline
\end{tabular}
\caption{\footnotesize Regular solutions of system \eqref{IMHD} and system \eqref{IMHD} with boundary conditions \eqref{BC}. 
In this notation, $SC^\infty\Omega_S\left[0\right]$ denotes the existence of at least one solution $\lr{\bol{B},P}$ with a symmetric ($S$) magnetic field $\bol{B}$ in the set of smooth functions $C^\infty$  
satisfying boundary conditions on a symmetric bounded domain $\Omega_S$. Such solution can be found in reference $[0]$.  
In a similar manner, $A$ stands for solution with an asymmetric magnetic field, and $\Omega_A$ for asymmetric bounded domain. 
When the symmetry property of the magnetic field is not known, it is omitted (e.g. class $b1$ $H^1\Omega_S$). 
For solutions of \eqref{IMHD} without boundary conditions, the listed solutions are valid in some bounded domain $\Omega$. 
References are specified only for the most nontrivial cases. The solution with the asterisk is the one 
derived in the present study. Explicit expressions can be found in tables \ref{tab2} and \ref{tab3} of the appendix. 
A solution is considered regular if at least of class $H^1$. 
}\label{tab1}
\end{center}
\end{table}  
The purpose of this study is to show that
class $p$ solutions of the type $SC^\infty\Omega_A$ exist (see the caption of table \ref{tab1} for details on this notation).
In particular, we will provide an explicit example of a 
translationally symmetric smooth magnetic field 
that solves system \eqref{IMHD} with boundary conditions \eqref{BC}
in an asymmetric ellipsoidal domain. 
	
The present paper is organized as follows. 
In section 2 we review the notion of symmetric solution, 
and discuss the relation between the symmetry of a solution and the topology of 
the bounding surface.
In section 3 we construct an example of class $p$ $SC^\infty\Omega_A$ solution.
Concluding remarks are given in section 4.

\section{Symmetry of Solutions}

Given a tensor $T$ and a vector field $\bol{\xi}$ in $\Omega$, 
$T$ is said symmetric with respect to $\bol{\xi}$ provided that
the Lie-derivative of $T$ along $\bol{\xi}$ identically vanishes, $\mf{L}_{\bol{\xi}}T=0$ in $\Omega$.
For the purpose of the present paper we are interested in the cases in which $T$ is either 
a twice covariant symmetric tensor $g=g_{ij}dx^i\otimes dx^j$, 
a vector field $\bol{B}=B^i\p_i$ or a function $P$. 
Then, the corresponding Lie-derivatives take the following expressions:
\begin{subequations}\label{SymSol}
\begin{align}
\mf{L}_{\bol{\xi}}g=&\lr{g_{jk}\frac{\p\xi^k}{\p x^i}+g_{ik}\frac{\p\xi^k}{\p x^j}+\frac{\p g_{ij}}{\p x^k}\xi^k}dx^i \otimes dx^j,\\
\mf{L}_{\bol{\xi}}\bol{B}=&\lr{\bol{\xi}\cdot\nabla}\bol{B}-\lr{\bol{B}\cdot\nabla}\bol{\xi},\\
\mf{L}_{\bol{\xi}}P=&\bol{\xi}\cdot\nabla P.
\end{align}
\end{subequations} 
In the context of magnetohydrodynamics, a symmetric solution 
is defined by the requirement that the components of the magnetic field $B^i$,
the pressure $P$, and the metric coefficients $g_{ij}$, $i,j=1,2,3$, are all independent of $x^3$,
where $x^3$ is the third coordinate of a curvilinear coordinate system $\lr{x^1,x^2,x^3}$:
\begin{equation}
\frac{\p B^i}{\p x^3}=0,~~~~\frac{\p P}{\p x^3}=0,~~~~\frac{\p g_{ij}}{\p x^3}=0,~~~~i,j=1,2,3.\label{SymSolMHD}
\end{equation} 
One can verify that \eqref{SymSolMHD} is equivalent to 
the vanishing of the Lie-derivatives in equation \eqref{SymSol} for $\bol{\xi}=\p_3$,
with $\p_3$ the tangent vector in the $x^3$ direction. 
The condition $\p g_{ij}/\p x^3=0$ restricts the class of admissible symmetries
to continuous Euclidean isometries (i.e. transformation that preserve the distance between
points in three dimensional Euclidean space, see \cite{Ed1,Ed2,SatoPPCF}).
These transformations are translations, rotations, or a combination of them. 
In these cases, the vector field $\bol{\xi}_E$ representing the direction of symmetry can be expressed as
\begin{equation}
\bol{\xi}_E=\bol{a}+\bol{b}\w\bol{x},
\end{equation}
where $\bol{a},\bol{b}\in\mathbb{R}^3$ are constant vector fields  
and $\bol{x}=\lr{x,y,z}^T$ is the position vector in $\mathbb{R}^3$.  
The vector field $\bol{a}$ is associated with translations, while
the vector field $\bol{b}$ generates rotations. 
Thus, in magnetohydrodynamics a magnetic field solving \eqref{IMHD} is asymmetric 
in a region $\Omega$ if there is no choice of $\bol{a}$ and $\bol{b}$
such that $\mf{L}_{\bol{\xi}_E}\bol{B}=\bol{0}$ in $\Omega$, except the trivial
case $\bol{a}=\bol{b}=\bol{0}$.
Notice that, for the purpose of the present study, we
shall consider a solution $\lr{\bol{B},P}$ 
to be symmetric provided that $\mf{L}_{\bol{\xi}_E}\bol{B}=\bol{0}$
for some non-trivial choice of $\bol{a}$ and $\bol{b}$
regardless of the symmetry of $P$. 

When solving the boundary value problem \eqref{IMHD}, \eqref{BC} it is customary to 
assume that the boundary $\p\Omega$ corresponds to a level set $\psi={\rm const.}$ of the 
stream (flux) function $\psi$, and that the pressure
is a function of $\psi$, $P=P\lr{\psi}$.   
Then, on $\p\Omega$, 
the unit outward normal $\bol{n}$ can be expressed as
\begin{equation}
\bol{n}=\frac{\nabla P}{\abs{\nabla P}}.
\end{equation}
Under these circumstances, the symmetry $\bol{\xi}_E=\p_3$ of a symmetric 
 solution \eqref{SymSolMHD} is inherited by the boundary $\p\Omega$ because
\begin{equation}
\mf{L}_{\bol{\xi}_E}\psi=\frac{d\psi}{d P}\bol{\xi}_E\cdot\nabla P=\frac{d\psi}{d P}\frac{\p P}{\p x^3}=0. 
\end{equation}
Furthermore, the existence of solutions of the boundary value problem \eqref{IMHD}, \eqref{BC} is restricted
to multiply connected domains (the domain $\Omega$ must have at least one hole).
This is because the hairy ball theorem \cite{Eisenberg} 
states that a tangent vector $\bol{B}$ to the boundary $P={\rm const.}$ of a simply connected domain $\Omega\subset\mathbb{R}^3$ 
must have a point where $\bol{B}=\bol{0}$, contradicting the fact that $\nabla P=\lr{\nabla\w\bol{B}}\w\bol{B}$  
is different from zero.

However, in a general setting the boundary $\p\Omega$ does not need to correspond to an
isobaric surface. Indeed, recalling the Clebsch parametrization \eqref{Clebsch}, 
in a small region $U$ centered on the boundary $\p\Omega$ we may express the magnetic field as
$\bol{B}=\nabla P\w\nabla\theta$ for some appropriate choice of the Clebsch potential $\theta$. 
Then, denoting by $\chi$ the 
function such that $\bol{n}=\nabla\chi/\abs{\nabla\chi}$ on $\p\Omega$, 
solutions $\chi$ of the equation $\bol{B}\cdot\bol{n}=0$ in $U$ 
take the form $\chi=\chi\lr{P,\theta}$. 
This result implies that a symmetric solution \eqref{SymSolMHD} 
may be enclosed by an asymmetric domain, provided that
the Clebsch potential $\theta$ does not exhibit the same symmetry $\bol{\xi}_E=\p_3$:
\begin{equation}
\mf{L}_{\bol{\xi}_E}\chi=\frac{\p\chi}{\p \theta}\bol{\xi}_E\cdot\nabla\theta=\frac{\p\chi}{\p\theta}\frac{\p\theta}{\p x^3}\neq 0. 
\end{equation} 
In the next section, we will construct an explicit example of a smooth translationally symmetric solution of system
\eqref{IMHD}, \eqref{BC} enclosed by an ellipsoidal asymmetric surface, i.e. a boundary that is not invariant
under continuous Euclidean isometries.

\section{A Symmetric Solution in an Asymmetric Ellipsoid}

\begin{figure}[h]
\hspace*{-0cm}\centering
\includegraphics[scale=0.22]{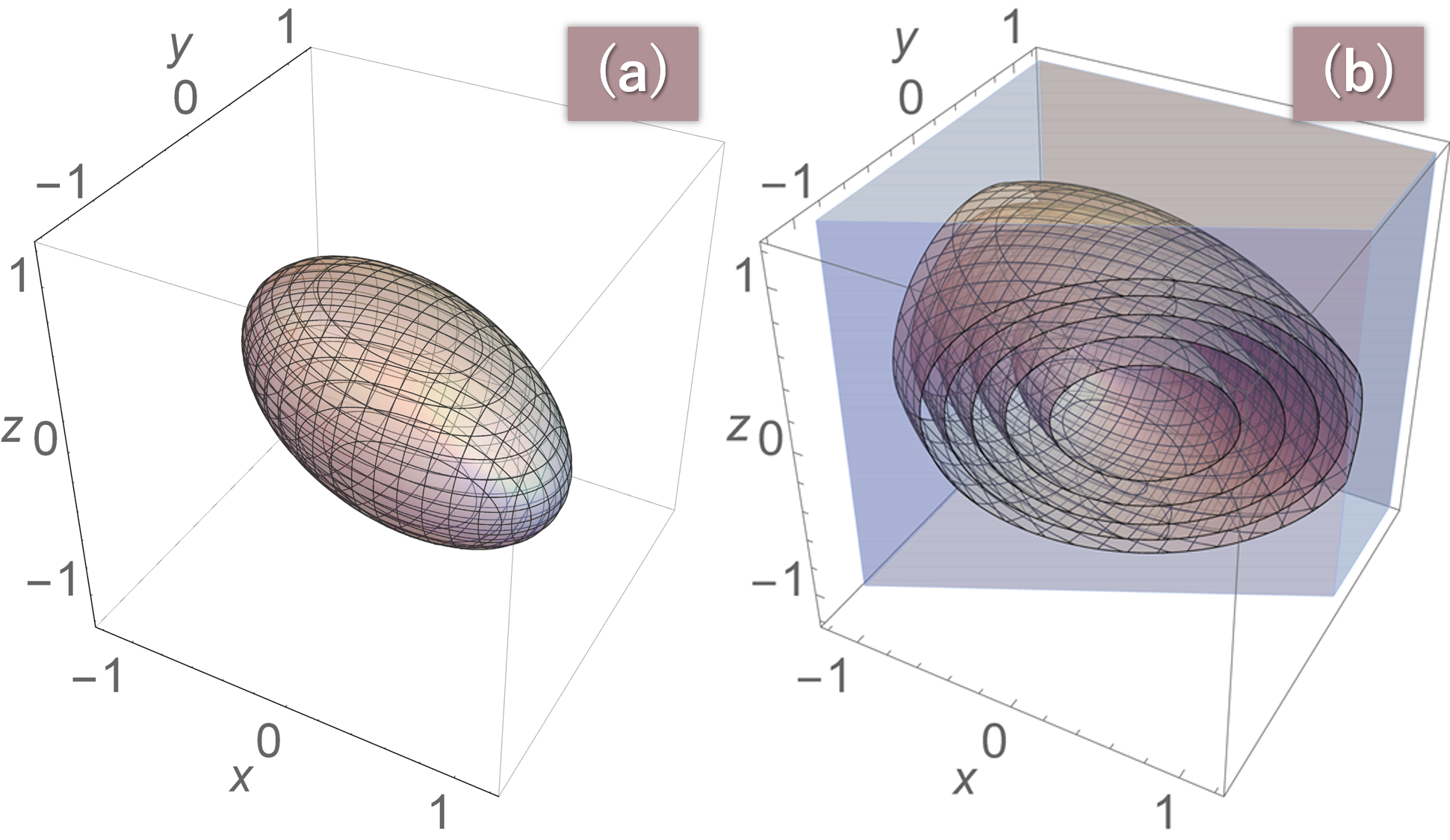}
\caption{\footnotesize (a) The level set $\psi=0.5$, with $\psi$ given by equation \eqref{Ell}. (b) Planar sections of the level sets 
$\psi=\left\{0.25,0.5,0.75,1.0,1.25\right\}$, with $\psi$ given by equation \eqref{Ell}.}
\label{fig1}
\end{figure}

An ellipsoid is a closed surface of the second order that
does not exhibit, in general, continuous Euclidean symmetries. 
It can be thought of as a deformation of the sphere, and as such
it represents one of the simplest asymmetric boundaries.
It should be noted that the corresponding domain $\Omega$ is simply connected. 
Examples of asymmetric ellipsoids can be obtained by the level sets of the function
\begin{equation}
\psi=\frac{1}{2}\lr{x^2+\frac{4}{3}xy+2 y^2+3 z^2}.\label{Ell}
\end{equation}
The contours of \eqref{Ell} are shown in figure \ref{fig1}.  

Let us show that the surfaces $\psi={\rm const.}$, with $\psi$ given by equation \eqref{Ell}, 
are not invariant under translations, rotations, or their combination. 
To see this, we must evaluate the Lie derivative
\begin{equation}\label{LEll}
\begin{split}
\mf{L}_{\bol{\xi}_E}\psi=&\frac{1}{2}\lr{\bol{a}+\bol{b}\w\bol{x}}\cdot\nabla\lr{x^2+\frac{4}{3}xy+2y^2+3z^2}\\=&
\lr{a_x+b_y z-b_z y}\lr{x+\frac{2}{3}y}\\&+\lr{a_y+b_z x-b_x z}\lr{\frac{2}{3}x+2y}\\&+3\lr{a_z+b_x y-b_y x}z. 
\end{split}
\end{equation}
Here, $\bol{a}=\lr{a_x,a_y,a_z}^T$ and $\bol{b}=\lr{b_x,b_y,b_z}^T$.
For $\psi$ in \eqref{Ell} to be symmetric, the Lie derivative \eqref{LEll} 
must vanish identically in the whole domain $\Omega$. 
First consider the line $\left\{x=y=0, \bol{x}\in\Omega\right\}$. 
Here, equation \eqref{LEll} reduces to $\mf{L}_{\bol{\xi}_E}\psi=3a_z z$, 
which implies $a_z=0$.  
Next consider the line $\left\{x=z=0, \bol{x}\in\Omega\right\}$.
Then, \eqref{LEll} gives the condition $a_x+3a_y=b_z y$, which implies $b_z=0$ and $a_x=-3a_y$.
Similarly, on the line $\left\{y=z=0, \bol{x}\in\Omega\right\}$ 
one obtains $a_x=a_y=0$. Now, equation \eqref{LEll} takes the form
\begin{equation}
\mf{L}_{\bol{\xi}_E}\psi=z\left[y\lr{b_x+\frac{2}{3}b_y}-x\lr{2b_y+\frac{2}{3}b_x}\right].\label{LEll2}
\end{equation}
Considering the planes $\left\{x=0, \bol{x}\in\Omega\right\}$ and
$\left\{y=0, \bol{x}\in\Omega\right\}$ separately, we conclude that 
also $b_x=b_y=0$ for the Lie derivative \eqref{LEll2} to vanish in $\Omega$. 
Hence, $\bol{\xi}_E=\bol{a}=\bol{b}=\bol{0}$. We have thus shown that the level sets $\psi={\rm const.}$
do not possess continuous Euclidean symmetries. 

Our next task is to construct a magnetic field $\bol{B}$ satisfying \eqref{IMHD}
with boundary conditions \eqref{BC} in a domain $\Omega$ whose boundary $\p\Omega$
is a level set of the function $\psi$ in equation \eqref{Ell}. 
To this end, we postulate the Clebsch parametrization $\bol{B}=\nabla\psi\w\nabla\theta$, 
and look for a Clebsch potential $\theta$ compatible with system \eqref{IMHD}, \eqref{BC}.
Notice that, by construction, $\nabla\cdot\bol{B}=0$ in $\Omega$ and $\bol{B}\cdot\bol{n}=0$ on $\p\Omega$.
Hence, we only need to choose $\theta$ so that the first equation in \eqref{IMHD} is satisfied
for some appropriate function $P$.  
In other words, it is sufficient to find a solution $\theta$ of the equation
\begin{equation}
\nabla\w\left\{\left[\nabla\w\lr{\nabla\psi\w\nabla\theta}\right]\w\lr{\nabla\psi\w\nabla\theta}\right\}=\bol{0}.\label{IMHDEll} 
\end{equation}
The idea is to accommodate $\bol{B}$ on $\p\Omega$ so that
it loops around planar sections of the ellipsoid. 
In such a configuration, we expect the electric current $\nabla\w\bol{B}$ to lie along 
the normal direction of the loops, and the resulting 
Lorentz force $\lr{\nabla\w\bol{B}}\w\bol{B}$ to be directed toward the center of the domain $\Omega$. 
Since in this construction $\bol{B}$ will be orthogonal to the plane normals, we expect the 
surfaces $\theta={\rm const.}$ to define planes in $\mathbb{R}^3$.
Therefore, we set
\begin{equation}
\theta=a x+b y+c z+d,~~~~a,b,c,d\in\mathbb{R}.\label{th}
\end{equation}
Substituting \eqref{Ell} and \eqref{th} into \eqref{IMHDEll}, we find
$a=2$, $b=-1$, $c=d=0$, and
\begin{equation}
\theta=2x-y.
\end{equation}
Then, the following identities can be verified:
\begin{subequations}
\begin{align}
\bol{B}=&3z\nabla\lr{x+2y}-\frac{7}{3}\lr{x+2y}\nabla z\label{BEll}\\=&z\lr{x+2y}\nabla\log\lr{\frac{\abs{x+2y}^3}{\abs{z}^{\frac{7}{3}}}},\\
\nabla\w\bol{B}=&\frac{16}{3}\nabla\lr{y-2x},\label{JEll}\\
\lr{\nabla\w\bol{B}}\w\bol{B}=&-\nabla\left[40 z^2+\frac{56}{9}\lr{x+2y}^2\right],\label{JxBEll}\\
\bol{B}^2=&\frac{49}{9}\lr{x+2y}^2+45 z^2,\label{B2Ell}\\
P=&-40 z^2-\frac{56}{9}\lr{x+2y}^2.\label{PEll}
\end{align}
\end{subequations}
Notice that the pressure $P$ is not a function of the stream function $\psi$ of \eqref{Ell}. 
Indeed,
\begin{equation}
\nabla P\w\nabla\psi=\frac{32}{3}\lr{2y^2-2x^2-3xy}z\nabla\log\lr{\frac{\abs{x+2y}}{\abs{z}^{7/3}}}.
\end{equation}

\begin{figure}[H]
\hspace*{-0cm}\centering
\includegraphics[scale=0.2]{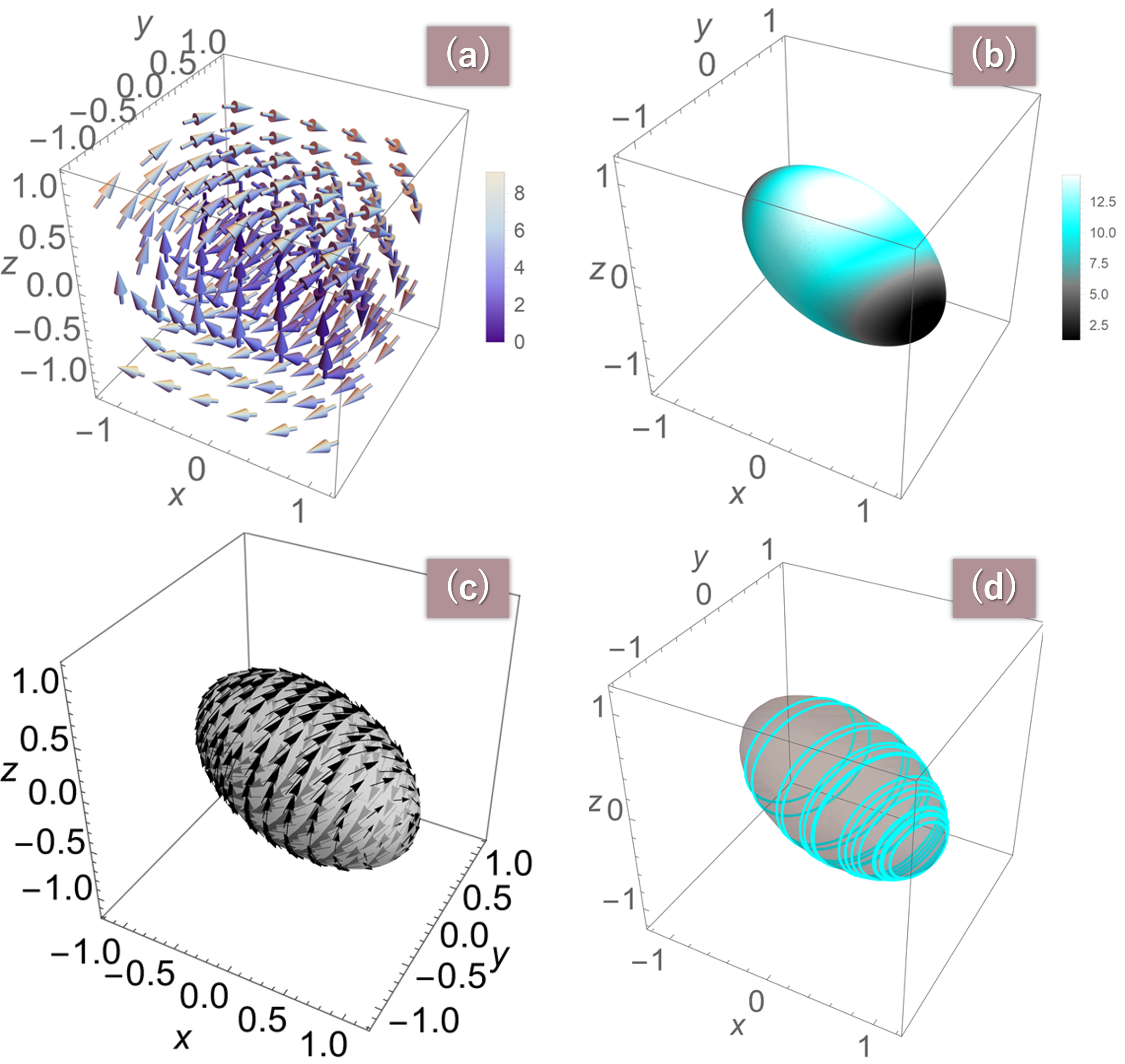}
\caption{\footnotesize (a) Vector plot of the magnetic field field \eqref{BEll}. (b) Plot of the magnetic field strength \eqref{B2Ell} on the isosurface (boundary) $\psi=0.5$ with $\psi$ given by \eqref{Ell}. (c) Vector plot of the magnetic field \eqref{BEll}  on the isosurface (boundary) $\psi=0.5$ with $\psi$ given by \eqref{Ell}. (d) A random set of magnetic field lines on the isosurface (boundary) $\psi=0.5$ with $\psi$ given by \eqref{Ell} and $\bol{B}$ given by \eqref{BEll}.} 
\label{fig2}
\end{figure}

\begin{figure}[H]
\hspace*{-0cm}\centering
\includegraphics[scale=0.2]{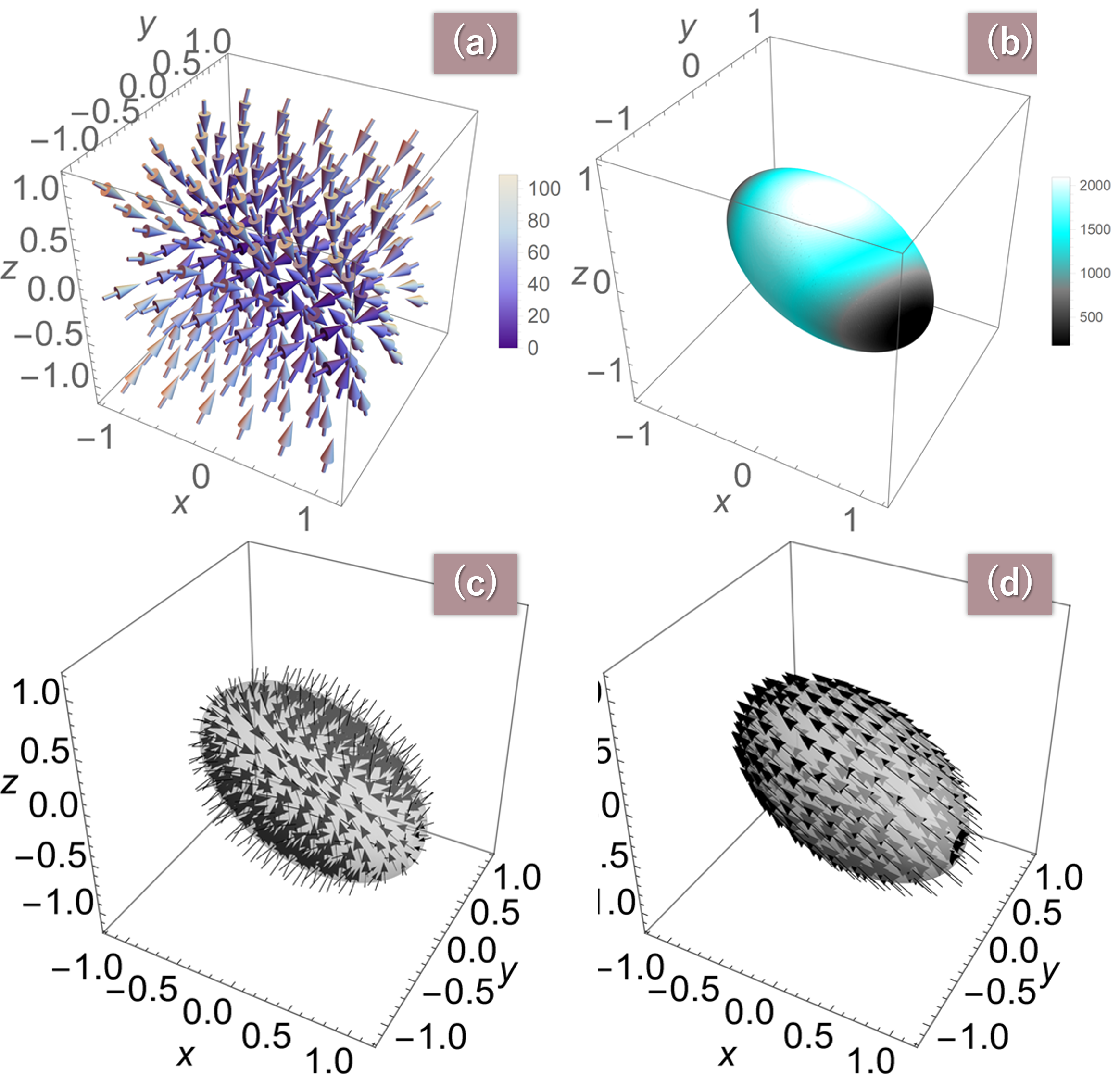}
\caption{\footnotesize (a) Vecor plot of the Lorentz force \eqref{JxBEll}. (b) Plot of the Lorentz force strength $\abs{\lr{\nabla\w\bol{B}}\w\bol{B}}^2$ on the isosurface (boundary) $\psi=0.5$ with $\psi$ given by \eqref{Ell} and the Lorentz force given by \eqref{JxBEll}. (c) Vector plot of the Lorentz force \eqref{JxBEll} on the isosurface (boundary) $\psi=0.5$ with $\psi$ given by \eqref{Ell}. (d) Vector plot of the electric current \eqref{JEll} on the isosurface (boundary) $\psi=0.5$ with $\psi$ given by \eqref{Ell}. Notice that the electric current is a constant vector field lying in the $\lr{x,y}$ plane.}
\label{fig3}
\end{figure}

\begin{figure}[h]
\hspace*{-0cm}\centering
\includegraphics[scale=0.22]{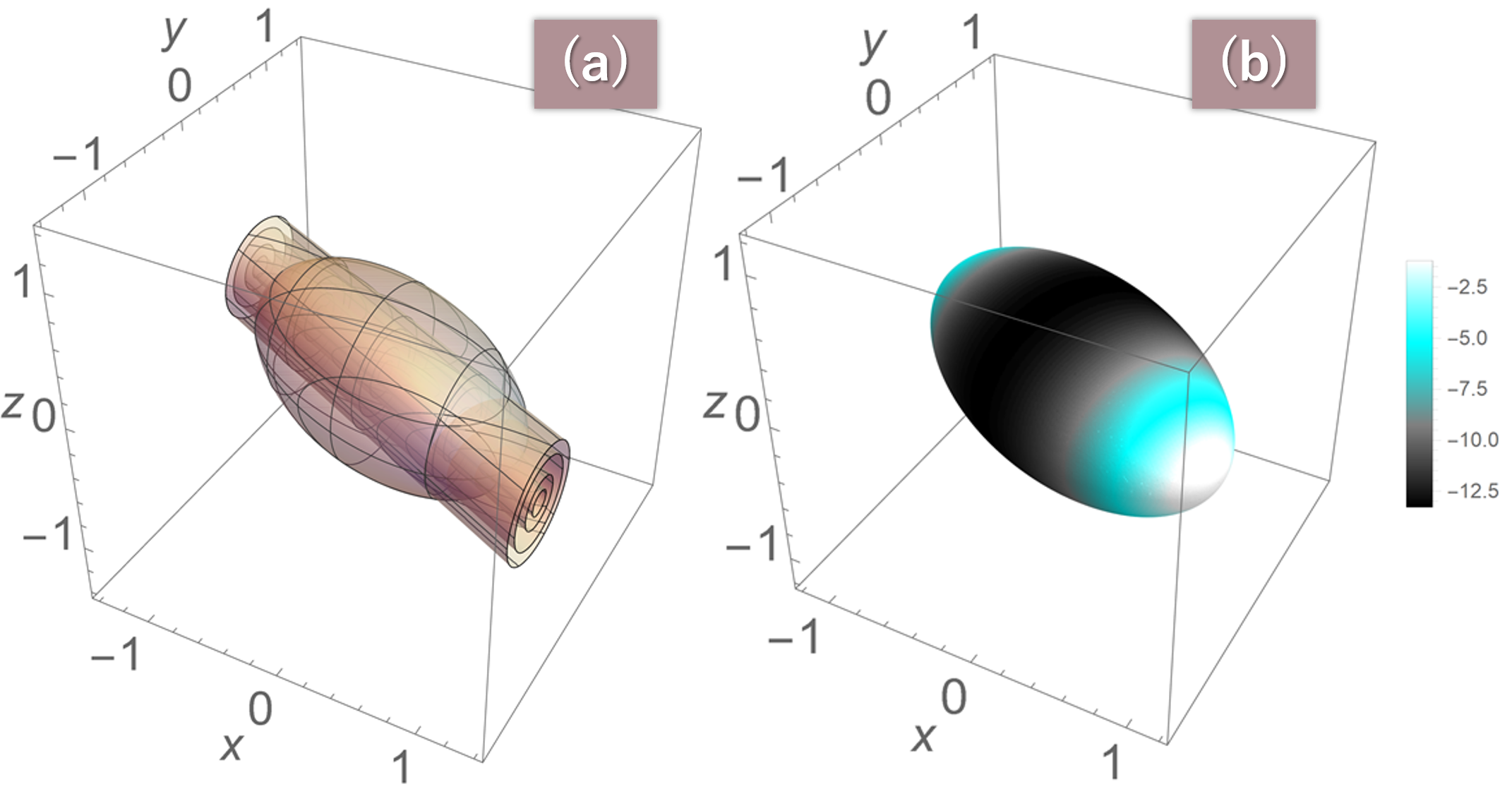}
\caption{\footnotesize (a) Isobaric surfaces $P=\left\{-0.25,-1,-3,-5\right\}$ intersecting the level set $\psi=0.5$ with $\psi$ given by \eqref{Ell}. (b) Plot of the pressure \eqref{PEll} on the level set $\psi=0.5$ with $\psi$ given by \eqref{Ell}.}
\label{fig4}
\end{figure}

A plot of the obtained magnetic field \eqref{BEll} is given in figure \ref{fig2} while
the Lorentz force \eqref{JxBEll} and the current \eqref{JEll} are shown in figure \ref{fig3}. 
The pressure \eqref{PEll} and the corresponding isobaric surfaces are depicted in figure \ref{fig4}. 

To conclude this section, let us verify that the obtained magnetic field \eqref{BEll} and pressure \eqref{PEll} 
possess a translational symmetry. To prove this, we must find a constant vector field $\bol{a}$ such that
\begin{equation}
\mf{L}_{\bol{a}}\bol{B}=\lr{\bol{a}\cdot\nabla}\bol{B}=\bol{0},~~~~\mf{L}_{\bol{a}}P=\bol{a}\cdot\nabla P=0.\label{LBP}
\end{equation}
Now observe that the second equation in \eqref{LBP} can be satisfyied by setting $a_x=2$, $a_y=-1$, and $a_z=0$. 
Using this expression for $\bol{a}$, the first equation in \eqref{LBP} becomes
\begin{equation}
\mf{L}_{\bol{a}}\bol{B}=-\left[\frac{7}{3}\nabla\lr{2x-y}\cdot\nabla\lr{x+2 y}\right]\nabla z=\bol{0}. 
\end{equation}
Hence, the solution \eqref{BEll}, \eqref{PEll} of the boundary value problem \eqref{IMHD}, \eqref{BC} 
in the region $\Omega$ enclosed by the level set $\psi={\rm constant}$, with $\psi$ given by \eqref{Ell}, 
has a translational symmetry in the direction $\bol{a}=\lr{2,-1,0}$ although $\p\Omega$
is asymmetric.

\section{Concluding Remarks}

In this study, we have constructed a smooth ideal magnetohydrodynamic equilibrium 
with a non-constant pressure that satisfies tangential boundary conditions
on an asymmetric ellipsoidal confinement vessel.
The obtained solution possesses a translational symmetry. 
One can verify that the same construction applies to asymmetric ellipsoids with different shapes.
This result shows that, 
by relaxing the standard assumption that the boundary corresponds to an isobaric surface, 
regular solutions of system \eqref{IMHD} with boundary conditions \eqref{BC} can be obtained in asymmetric bounded domains.

While the obtained solution solves an outstanding mathematical problem, 
it also suggests the possibility of increasing the degree of freedom in the design of    
plasma confinement devices by relaxing the usual boundary conditions,  
which topologically constrain the region $\Omega$ to a topological torus.   

We remark that the existence of regular solutions 
of the types $b2$ $SC^\infty\Omega_A$, $b2$ $AC^\infty\Omega_A$, and $p$ $AC^\infty\Omega_A$, 
i.e. Beltrami field solutions with non-constant proportionality factor in asymmetric domains, and
asymmetric non-vanishing pressure gradient solutions in asymmetric domains, 
remains an open question that deserves further investigation.  
Similar considerations hold for the existence
of asymmetric solutions in symmetric domains such as $b2$ $AC^\infty\Omega_S$ and $p$ $AC^\infty\Omega_S$, 
as well as the classes $h$ $AC^{\infty}\Omega_S$, $h$ $SC^{\infty}\Omega_A$, $h$ $AC^{\infty}\Omega_A$, $b2$ $SC^{\infty}\Omega_S$, $b2$ $AC^{\infty}\Omega_S$ which are, at present, unexplored.
Finally, the symmetry properties of solutions belonging to class $b1$ $H^1\Omega_S$ and $b1$ $H^1\Omega_A$ are also unclear.


\begin{appendix}

\section{Explicit Expressions of Solutions}
Tables \ref{tab2} and \ref{tab3} provide
explicit expressions for solutions of system \eqref{IMHD}
and system \eqref{IMHD} with boundary conditions \eqref{BC} respectively. 

\begin{table}[H]
\begin{center}
\begin{tabular}{ |c|c| }
\hline
Class & Solutions of \eqref{IMHD} \\
\hline
$h$ $SC^{\infty}$ & $\bol{B}=\nabla x$\\
 & $P=0$\\
& $\bol{\xi_E}=\lr{a_x,a_y-b_x z,a_z+b_x y}^T$\\
\hline 
$h$ $AC^{\infty}$ & $\bol{B}=\nabla\lr{e^x\sin{y}+e^{z}\cos{x}}$\\
& $P=0$\\
\hline
$b1$ $SC^{\infty}$ & $\bol{B}=e^x\sin\lr{y+z}\nabla y$ \\
& $~~~~~~~~-e^x\cos\lr{y+z}\nabla x$ \\
 & $P=0$\\
& $\bol{\xi}_E=\lr{0,-a_z,a_z}^T$\\
\hline
$b1$ $AC^{\infty}$ & $\bol{B}=e^y\sin\lr{x+z}\nabla z$ \\
& $~~~~~~~~-e^x\cos\lr{y+z}\nabla x$\\
& $~~~~~~~~~+e^x\sin\lr{y+z}\nabla y$\\
& $~~~~~~~~~-e^y\cos\lr{x+z}\nabla y$\\
& $P=0$\\
\hline
$b2$ $SC^{\infty}$ & $\bol{B}=\sin\lr{z^2}\nabla x+\cos\lr{z^2}\nabla y$\\
 & $P=0$\\
& $\bol{\xi}_E=\lr{a_x,a_y,0}^T$\\
\hline
$b2$ $AC^{\infty}$ & $\bol{B}=e^x\sin\lr{y+z^2}\nabla y$\\
& $~~~~~~~~-e^x\cos\lr{y+z^2}\nabla x$ \\
& $P=0$\\
\hline
$p$ $SC^{\infty}$ & $\bol{B}=3z\nabla\lr{x+y}-\lr{x+y}\nabla z$ \\
 & $P=-2\lr{x+y}^2-12z^2$\\
& $\bol{\xi}_E=\lr{a_x,-a_x,0}^T$\\
\hline
$p$ $AC^{\infty}$ & $\bol{B}=\lr{x+e^{-z}}\nabla x-y\nabla y+\nabla z$\\
& $P=-e^{-z}\lr{x+\frac{e^{-z}}{2}}$\\
\hline
\end{tabular}
\caption{\footnotesize Expressions of solutions to system \eqref{IMHD}. The pressure $P$ and, when available,
the continuous Euclidean symmetry $\bol{\xi}_E$ of the magnetic field are also specified.
}\label{tab2}
\end{center}
\end{table} 

\begin{table}
\begin{center}
\begin{tabular}{ |c|c|c| }
\hline
Class &  Solutions of \eqref{IMHD}, \eqref{BC}\\
\hline
$h$ $SC^{\infty}\Omega_S$ & $\bol{B}=\frac{x}{x^2+y^2}\nabla y-\frac{y}{x^2+y^2}\nabla x$ \\
& $P=0$\\
& $\bol{\xi}_E=\lr{-b_z y, b_z x,a_z}^T$\\
& $\psi=\left[1-\sqrt{x^2+y^2}\right]^2+z^2$\\
& $\bol{\chi}_E=\lr{-b_z y,b_z x,0}^T$\\
\hline 
$p$ $SC^{\infty}\Omega_S$ & $\bol{B}=3z\nabla\lr{x+y}-\lr{x+y}\nabla z$\\
& $P=-2\lr{x+y}^2-12z^2$\\
& $\bol{\xi}_E=\lr{a_x,-a_x,0}^T$\\
& $\psi=\frac{1}{2}\lr{x^2+y^2+3z^2}$\\
& $\bol{\chi}_E=\lr{-b_z y,b_z x,0}^T$\\
\hline
$p$ $SC^{\infty}\Omega_A$ & $\bol{B}=3z\nabla\lr{x+2y}-\frac{7}{3}\lr{x+2y}\nabla z$\\
& $P=-40 z^2-\frac{56}{9}\lr{x+2y}^2$\\
& $\bol{\xi}_E=\lr{-2a_y,a_y,0}^T$\\
& $\psi=\frac{1}{2}\lr{x^2+\frac{4}{3}xy+2y^2+3z^2}$\\
\hline
\end{tabular}
\caption{\footnotesize Expressions of solutions to system \eqref{IMHD} with boundary conditions \eqref{BC}. 
Here, $\psi$ is the function whose level set generates the bounding surface $\p\Omega$. 
The pressure $P$ and, when available, the continuous Euclidean symmetry $\bol{\xi}_E$ of the magnetic field   
and the continuous Euclidean symmetry of the boundary $\bol{\chi}_E$ are also specified. 
Notice that the symmetry of the magnetic field $\bol{\xi_E}$ does not necessarily coincide with 
the symmetry of the boundary $\bol{\chi}_E$.}\label{tab3}
\end{center}
\end{table}

\end{appendix}

\section*{Acknowledgment}
The research of NS was partially supported by JSPS KAKENHI
Grant No. 17H01177.
The author is grateful to Professor Z. Yoshida for useful discussion. 

\section*{Data Availability}
The data that support the findings of this study are available from the corresponding author upon reasonable request.

\end{document}